\documentclass[12pt]{article}

\usepackage{amsmath,amssymb,amsfonts,graphicx,color}

\newcommand\pubnumber{SNSN-323-63}
\newcommand\pubdate{\today}

\def\napoli{Leung Center for Cosmology and Particle Astrophysics\\National Taiwan University}

\def\Title#1{\begin{center} {\Large #1 } \end{center}}
\def\Author#1{\begin{center}{ \sc #1} \end{center}}
\def\Address#1{\begin{center}{ \it #1} \end{center}}

\newcommand\pubblock{\rightline{\begin{tabular}{l} \pubnumber\\
         \pubdate  \end{tabular}}}
\newenvironment{Abstract}{\begin{quotation}  }{\end{quotation}}
\newenvironment{Presented}{\begin{quotation} \begin{center} 
             PRESENTED AT\end{center}\bigskip 
      \begin{center}\begin{large}}{\end{large}\end{center} \end{quotation}}
\def\Acknowledgements{\bigskip  \bigskip \begin{center} \begin{large}
             \bf ACKNOWLEDGEMENTS \end{large}\end{center}}

\def\beq{\begin{equation}}
\def\eeq#1{\label{#1}\end{equation}}
\def\eeqn{\end{equation}}

\def\beqa{\begin{eqnarray}}
\def\eeqa#1{\label{#1}\end{eqnarray}}
\def\eeqan{\end{eqnarray}}

\let\bar=\overbar

\def\Dslash{\not{\hbox{\kern-4pt $D$}}}
\def\dslash{\not{\hbox{\kern-2pt $\del$}}}

\def\msb{{\bar{\ssstyle M \kern -1pt S}}}

\begin{document}
\begin{titlepage}
\pubblock

\vfill
\Title{Remarks on Realizing Inflation in String and Particle Physics Models}
\vfill
\Author{Sean Downes}
\Address{\napoli}
\vfill
\begin{Abstract}
The recent measurements of cosmological parameters by the Planck collaboration favors inflationary models with a redshifted spectrum and a very low tensor-to-scalar ratio. Two well studied scenarios in the particle physics/string theory literature --- inflection point and Starobinsky inflation ---  are in good agreement with these data. In these proceedings, we report on general studies of these two scenarios. We discuss similarities in their structure which, we argue, arise from a common mechanism: degenerate critical point of the potential. In particular, we emphasize their ability to support a possible primordial explanation for the anomalous suppression of the power spectrum at large scales.
\end{Abstract}
\vfill
\begin{Presented}
The 10th International Symposium on Cosmology and Particle Astrophysics (CosPA2013)\\Honolulu, HI, USA, November 12–15, 2013

\end{Presented}
\vfill
\end{titlepage}
\def\thefootnote{\fnsymbol{footnote}}
\setcounter{footnote}{0}
%


\section{Motivation}
\subsection{Introduction}
Inflation uses a scalar field coupled to a Friedmann-Lema\^itre-Robertson-Walker (FLRW) metric to mimic de Sitter background for spacetime \cite{Guth:1980zm,Starobinsky:1980te,Albrecht:1982wi,Linde:1981mu}. An early burst of exponential expansion resolves many technical deficiencies in the na\"ive Big Bang scenario, while predicting nearly scale-invariant perturbations \cite{Mukhanov:1981xt} to the curvature of the FLRW spacetime necessary for structure formation in the universe. These same perturbations were predicted to jostle the primordial plasma of nuclei and electrons, leaving an imprint on the cosmic microwave background (CMB). With the advent of precision cosmological experiments including the recent results from Planck Collaboration \cite{Ade:2013uln}, the statistics of CMB temperature fluctuations have become precise enough to assess the viability of inflationary models.

One notable result is Planck's definitive observation of a deviation of the spectral index from unity. This strongly suggests that inflation is a dynamical phenomena, as opposed to de Sitter period of false vacuum inflation. The strongly Gaussian statistics of the temperature fluctuations sharpen this argument for a period ``slow-roll'' inflation, where the dynamics are essentially determined by the scalar potential. The further non-observation of CMB polarization signatures associated with tensor modes during inflation further reduce the class of possible inflationary models. This guidance feeds directly into attempts to embed inflation into larger models in particle physics or String Theory. The work we report on here examines generic dynamical properties and observational signatures of the onset of inflation.

Taken together, the observation a redshifted spectrum with the (as-yet) unobservably small production of tensor modes and nongaussianity suggests a rather vanilla slow-roll scenario with an extremely small velocity. For the case of a single field, both the tensor-to-scalar ratio $r$ and (local) $f_{\rm NL}$ are effectively proportional to the square the field velocity. These features are in tension the simple monomial potentials, but are generic to potentials which admit degenerate critical points at nonzero vacuum energy. Two such scenarios include inflection point inflation and the Starobinsky potential. The former involves a doubly degenerate critical point somewhere above the reheating vacuum, while the latter involves an infinitely degenerate critical point at infinite distance in field space. Curiously, both scenarios are rampant in phenomenological studies. While ease of construction is not an argument for correctness, it does afford a deeper analysis, which is certainly warranted by the favorable observations. 
\subsection{E-foldings as a tuning}

In traditional, large-field models of slow-roll inflation \cite{Linde:1983gd}, the total number of e-foldings was determined by the initial field vacuum expectation value (VEV). Typical arguments involve Planck-scale physics kicking the field out to such large VEV, which can then be self-sustaining via quantum effects. In practice this approach can be difficult to realize. Assuming one can build a consistent effective field theory (EFT) to describe the physics, Planck-scale corrections to the potential are generally expected to violate the slow-roll conditions, thereby ending inflation. While techniques to circumvent this problem can be arranged, the fact remains that such a scenario applies Planck-scale physics to set the initial conditions for inflation and effectively ignore it for the subsequent evolution. Without a clear mechanism, correlated to other observable physics, such approach essentially amounts to a fine-tuning of $N_e$. This is a general problem in inflationary cosmology.

Another approach to this tuning is to embed it into the couplings within the potential. Here the problem is more precise and amenable to standard techniques of effective field theory. Since both the inflection point and Starobinsky scenarios arise from a similar mathematical phenomena --- degenerate critical points --- it is perhaps unsurprising their physics admit a similar parametrization. In both cases, the number of e-foldings scales parametrically with inverse powers of a small coupling. While the choice of initial conditions can be further tuned to arrive at a smaller $N_e$, in both cases the dynamics enforces a maximum value. For the case of inflection point inflation in particular, this feature --- that overshooting the inflection point is entirely avoidable --- is highly nontrivial. We shall describe these dynamics in detail below.

If the expected value of $N_e$ is indeed a fine-tuning, naturalness --- now defined in terms of the EFT couplings --- suggests the minimum possible value of $N_e$ is also the most likely value. Observational implications of this suggest a large-scale cutoff in correlation of primordial curvature perturbation; all modes which exited the horizon subsequently re-enter. Curiously, the loss of correlation in the temperature power spectrum at large scales has persisted through WMAP \cite{Bennett:2003bz} to Planck. In what follows we shall also explore the generic features associated the possibility that this large-scale anomaly is caused by the onset of inflation. This subject has a rich literature, including foundational works like \cite{Polarski:1992dq,Contaldi:2003zv}, and a number of phenomenological analyses, including \cite{Cline:2003ve,Jain:2008dw,Ramirez:2012gt,Dudas:2012vv,Biswas:2013dry,Downes:2012gu,Cicoli:2013oba,Pedro:2013pba,Bousso:2013uia}.

\subsection{Slow-Roll as Dynamical Attractor}
The familiar field equation for inflation,
\begin{equation}\ddot{\phi}+3H\dot{\phi}+V_{\phi} = 0,\end{equation}
can be rewritten using $N=\log a$ to parametrize time,
\begin{equation}\label{eqn:FEQ}\phi^{\prime\prime}=\frac{1}{2}\left(\phi^{\prime}+\sqrt{6}\right)\left(\phi^{\prime}-\sqrt{6}\right)\left(\phi^{\prime}+\frac{V_{\phi}}{V}\right).\end{equation}
here primes denote derivatives with respect to $N$. There are two singular solutions to this differential equation, $\phi^{\prime}\pm \sqrt{6}$. These represent kinetic energy dominated solutions. It is an artifact of the parametrization that $|\phi^{\prime}|<\sqrt{6}$. A third singular solution is possible, 
\begin{equation}\label{eqn:SR}\phi_{\rm SR}^{\prime} = -V_{\phi}/V,\end{equation}
provided that 
\begin{equation}\label{eqn:exact}\phi_{\rm}^{\prime\prime} = 0.\end{equation}
This is true, for example, if $V = e^{-\kappa\phi}$. In this case \eqref{eqn:FEQ} becomes a first order differential equation in $\phi^{\prime}$. If $|\kappa|<\sqrt{6}|$, all nonsingular trajectories attract (up to a constant) to $\phi_{\rm SR}$ \cite{Downes:2012xb}. Importantly, this behavior persists when \eqref{eqn:exact} is only approximately satisfied. Instead we find,
\begin{equation}\label{eqn:inexact}\phi_{\rm}^{\prime\prime} = \frac{V_{\phi\phi}}{V} - \left(\frac{V_{\phi}}{V}\right)^2\ll 1,\end{equation}
from which one can read off the traditional slow-roll parameters. \eqref{eqn:inexact} is a sufficient condition for inflation, and is our operational definition of ``slow-roll''. Integrating \eqref{eqn:SR} allows us to approximate the number of e-foldings,
\begin{equation}\label{eqn:Ne}N = -\int d\phi\; \frac{V}{V_{\phi}}.\end{equation}
Finally, let us note that there are two dynamical quantities which can be used to parametrize a bought of inflation. The first is $\Xi=\phi^{\prime\prime}/\phi^{\prime}$, which determines how close $\phi$ is to $\phi_{\rm SR}$. The second is $\frac{1}{2}\phi^{\prime 2}$, which determines just how slow slow-roll is. Note that we implicitly assume that \eqref{eqn:inexact} is satisfied. These are essentially the ``generalized slow-roll'' parameters of \cite{Dvorkin:2009ne}.

\section{Inflection Point Models}

\subsection{Inflection Points as Dynamical Fixed Points}

Inflection point inflation satisfies \eqref{eqn:inexact} with a (doubly) degenerate critical point of $V$, $\phi_{\star}$ --- \textit{i.e.}, $V_{\phi}$ and $V_{\phi\phi}$ vanish there. For a single scalar field, any smooth $V$ with such an ``inflection point'' can always be Taylor expanded,
\begin{equation}V\approx \Lambda + \alpha (\phi-\phi_{\star})^3 + \mathcal{O}(\phi^4).\end{equation}

This strategy is conceptually easy to deploy in phenomenological models: one need only tune a parameter so that
$V_{\phi\phi}\approx 0$ near a critical point. As such it has been used across the board, from particle physics models like the MSSM \cite{Allahverdi:2006iq}, to KKLT \cite{Kachru:2003aw} inspired supergravity constructions. Curiously, it appears to fail in Higgs inflation \cite{Hamada:2013mya}, owing to the running of the quartic self coupling.

Since inflation occurs near $\phi_{\star}$, higher order terms contribute little to the dynamics; $\Lambda$ the is dominant contribution to $V$.  At it is, this potential can generate arbitrarily large $N_e$. This fact --- that overshoot is not a generic phenomena --- was studied in \cite{Allahverdi:2008bt} and explicity demonstrated in \cite{Itzhaki:2008hs}. This fixed-point behavior was studied in detail in \cite{Downes:2012xb}, which also examined arbitrary initial conditions and the basin of attraction phase space.

Importantly, Itzhaki and Kovetz demonstrated that this attractor behavior persists with relevant deformations, linear and quadratic in $\phi$. By suitably shifting the origin of field space we can parametrize these deformations as
\begin{equation}\label{eqn:generalIPI}V\approx \Lambda + \lambda\phi+\alpha \phi^3 + \mathcal{O}(\phi^4).\end{equation}
Applying \eqref{eqn:Ne} then, one finds
\begin{equation}N_e \approx \frac{\pi}{2\sqrt{3\alpha\lambda}}.\end{equation}
For $\alpha$ near the scale of inflation, \textit{i.e.} $H^2 \approx \alpha^4/M_P^2$, we see that the smallness of $\lambda$ controls $N_e$. Note that we are implicitly assuming $\lambda$ and $\alpha$ have the same sign, otherwise a local minima would slightly complicate the analysis, but for small $|\lambda|$ the results are qualitatively similar \cite{Linde:2007jn}.

\subsection{The Onset of Inflation}

The fixed point behavior of the inflection point manifests itself as a transition from a large-field ``chaotic'' inflation to small-field inflection point inflation. The concavity of the potential --- and subsequently the inflaton trajectory --- flips. This transition induces a brief departure from slow-roll at the onset of inflection point inflation. Near the boundary of the basin of attraction, this transition can be extremely sharp --- accelerated expansion may even temporarily stop. The effect on the primordial curvature perturbation would be a sharp reduction in its power spectrum. 

To model this behavior, consider the potential studied in \cite{Downes:2012xb},
\begin{equation}\label{eqn:sf}\frac{1}{4}\phi^4 + \alpha\phi^3 + \lambda\phi + \left(\frac{27}{4}\alpha^4 + 3\alpha\lambda\right)\alpha^4.\end{equation}


The power spectrum of temperature fluctuations in the CMB has a sharp reduction in power at the largest scales. The power spectrum is expected to be flat at low angular moments, $\ell\lesssim50$, since the baryon acoustic oscillations do not appreciably affect such scales. This is contrary to what is observed. The anomalously low power in the temperature power spectrum of the CMB, observed in COBE and WMAP has persisted in Planck data release. While its statistical significance is somewhat muted by cosmic variance and concerns about the integrated Sachs-Wolfe effect \cite{Finelli:2005zc}, the anomaly's persistence and its possible explanation in terms of the onset of inflation warrants a theoretical investigation.

\begin{figure}[h!]\centering
\includegraphics[width=0.45\textwidth]{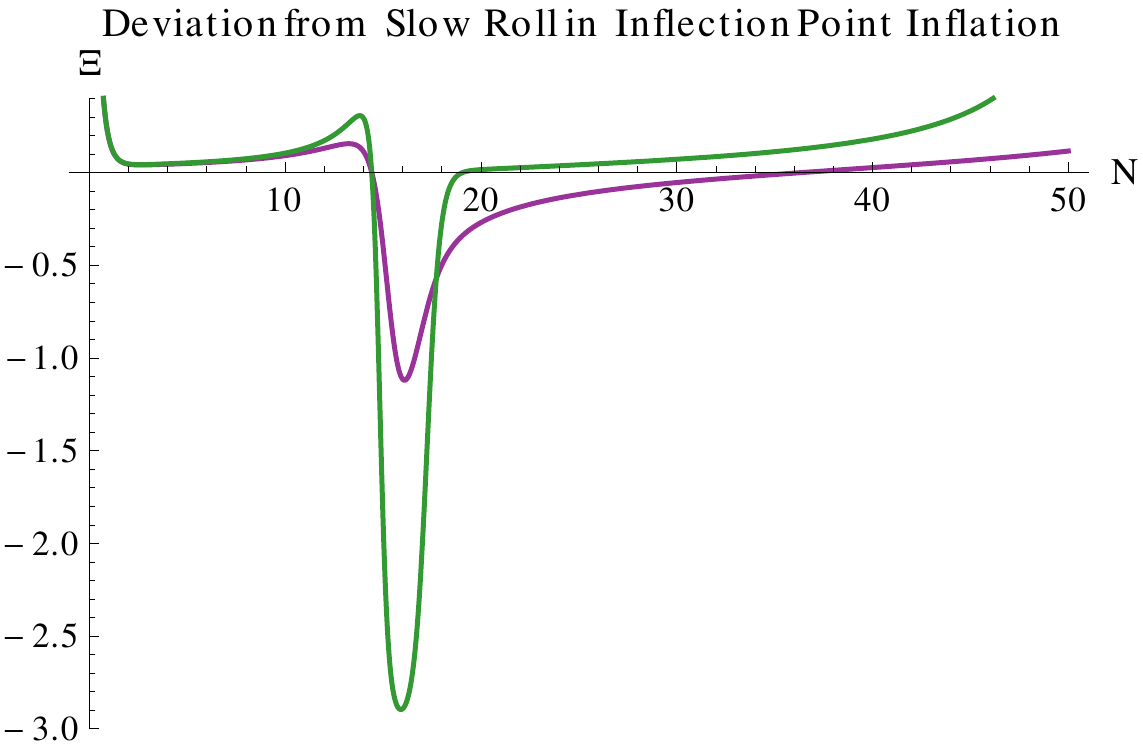}~\includegraphics[width=0.45\textwidth]{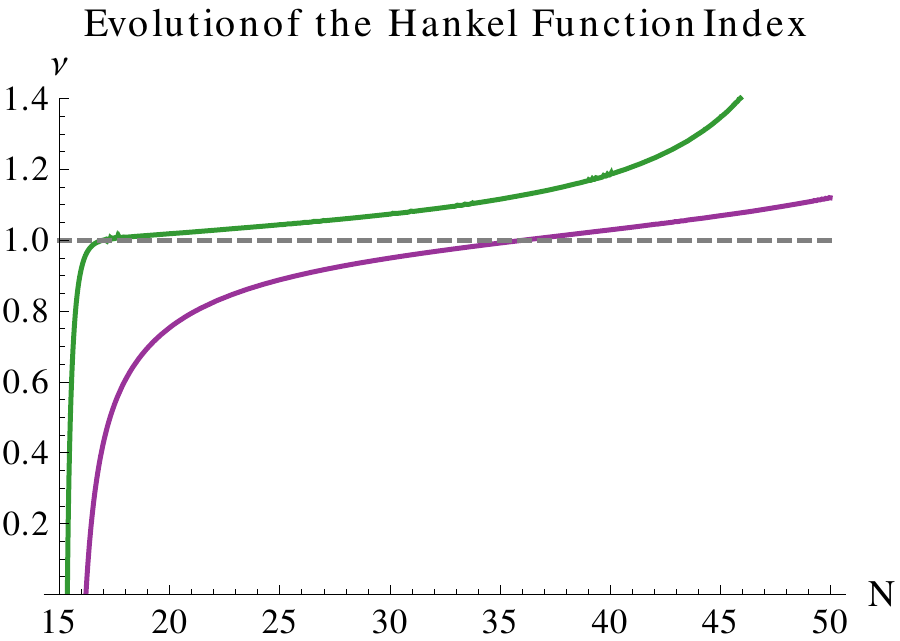}
\label{fig:ipichi}
\end{figure}

In \cite{Downes:2012gu}, these effects were studied using the brief injection of kinetic energy from the transition to inflection point inflation. Note that the slow-roll conditions are explicitly violated during the transition, which explains why one may attain an blue-to-redshifted``asymmetric'' (in terms of e-foldings) evolution of the power spectrum. In Fig.~\ref{fig:ipichi} we give examples of soft and sharp transitions. This gives rise to a suppression of the lowest few angular moments of the power spectrum \cite{Downes:2012gu}. 

\section{Starobinsky-like Models}
\subsection{Exponentials in the Scalar Potential}
Inflation in the Starobinsky scenario \cite{Starobinsky:1980te} (and slight generalizations thereof) involves a scalar field with a long plateau generated by the interplay of constant vacuum energy with an exponential,
\begin{equation}\label{eqn:staro}V(\phi) = \Lambda + A e^{-a\phi},\end{equation}
for some positive, real constants $\Lambda,\alpha$ and $\kappa$.
The potential \eqref{eqn:staro} has an infinitely degenerate critical point at infinite field values. More generally one might --- and in some cases, should --- consider more general potentials with many exponentials,
\begin{equation}V(\phi) = \Lambda + A e^{-a\phi} + B e^{b\phi}+\cdots.\end{equation}
If such additional terms contribute more \textit{real} critical points at finite VEV, one is led back to the KKLT version of the inflection point scenario. Of particular interest is the sign of $b$. If $b$ is positive, this removes the critical point at infinity and gives a finite length to the plateau. A length which can induce sufficient inflation requires on $B$ being a small number. Provided such a model is under parametric control, this can give rise to an upper bound on the number of e-foldings. As with the $\lambda$ coupling in \eqref{eqn:generalIPI}, this condition ports the fine-tuning required to attain $N_e\sim 50$ to the potential couplings.

Given that exponential terms in the scalar potential abound in string/supergravity models, this scenario has been well-studied \cite{Cicoli:2008gp,Cicoli:2011ct,Ellis:2013xoa,Farakos:2013cqa}. For our present purposed, we shall use Fibre Inflation \cite{Cicoli:2008gp} as an avatar of the Starobinsky scenario.
\subsection{Fibre Inflation}

The Fibre Inflation model exists within the LARGE volume scenario (LVS) framework \cite{Cicoli:2008va}. As such, the compactification manifold necessarily admits a limit wherein the size of SUSY cycle can be made parametrically smaller than the overall volume of the Calabi-Yau three-fold ($CY_3$), which is taken to be very large. This hierarchy introduces a scalar which stabilizes the potential during inflation. Operationally, any terms generated by Planck scale effects will be suppressed by inverse powers of the $CY_3$ volume. Thus, the $\eta$-problem is avoided.


The scalar potential for $\tau$ is generated by string loop corrections. The precise calculation of the scalar potential for a specific $CY_3$ remains a challenge, but analysis of Orientifold calculations \cite{Berg:2005ja} suggest terms like
\begin{equation}\label{eqn:fibs}V = \frac{A}{\tau} + \frac{B}{\mathcal{V}\sqrt{\tau}}+\frac{C\tau}{\mathcal{V}^2}+\cdots.\end{equation}
are generically present. The ellipses represent other possible terms that scale as
\begin{equation}C_{n}\frac{\tau^{n+1}}{\mathcal{V}^{2(1+n/3)}}.\end{equation}
Canonically normalizing the field $\tau$, 
$$\tau \rightarrow \mathcal{V}^{2/3}e^{2\phi\sqrt{3}},$$
and appropriate redefining the parameters as in \cite{Cicoli:2013oba}, the scalar potential \eqref{eqn:fibs} becomes,
%
%
%
\begin{equation}\label{eqn:fibre}V \simeq \frac{m_{\phi}^2}{4}\left[\left(1+\frac{2}{3}\delta\right) e^{-4\phi\sqrt{3}} - 4\left(1+\frac{\delta}{6}\right)e^{-\phi/\sqrt{3}} + \frac{\delta}{1+n}e^{2(1+n)\phi/\sqrt{3}} + 3-\frac{\delta}{1+n}\right].\end{equation}
%
%
%

Here we have included the two ``lowest'' terms, with the third left general as in \cite{Cicoli:2013oba}. $m_{\phi}$ represents the inflaton mass about its vacuum at the origin, it also sets the scale for inflation. For $n\geq1$, $\delta$ determines the number of e-foldings as discussed above. This is the case we shall consider below. For $n=1$, this is the original model of \cite{Cicoli:2008gp}. For positive $n$, this term represents the exponential ``wall'' for the inflaton at large VEV.

Assuming the field starts above the plateau, the number of e-foldings is determined by $\delta$. For the case of $n=1$, $\delta\approx10^{-7}$ gives around sixty e-foldings of inflation. For general $n$,
\begin{equation}N_e \propto \frac{9}{4}\delta^{-\frac{1}{3+2n}}.\end{equation} 

An attractive feature of the Fibre Inflation scenario is that such a small value of $\delta$ is natural from the EFT standpoint. $\delta$ depends parametrically on the string coupling, $g_s$ which is the EFT expansion parameter.
Indeed, 
\begin{equation}\delta = \mathcal{O}(g_s^{4(1+n/3)}).\end{equation}
Sixty e-foldings of inflation, therefore, require a value of $g_s \sim \mathcal{O}(10^{-2})$ \cite{Cicoli:2013oba}. We now study how the onset of inflation may induce a large scale suppression of the power spectrum.

\begin{figure}[h!]\centering
\includegraphics[width=0.45\textwidth]{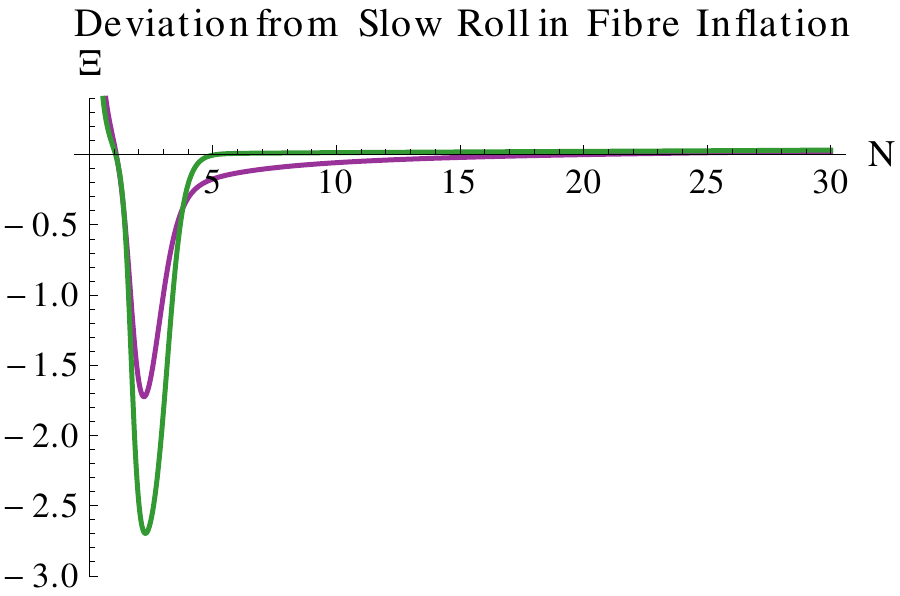}~\includegraphics[width=0.45\textwidth]{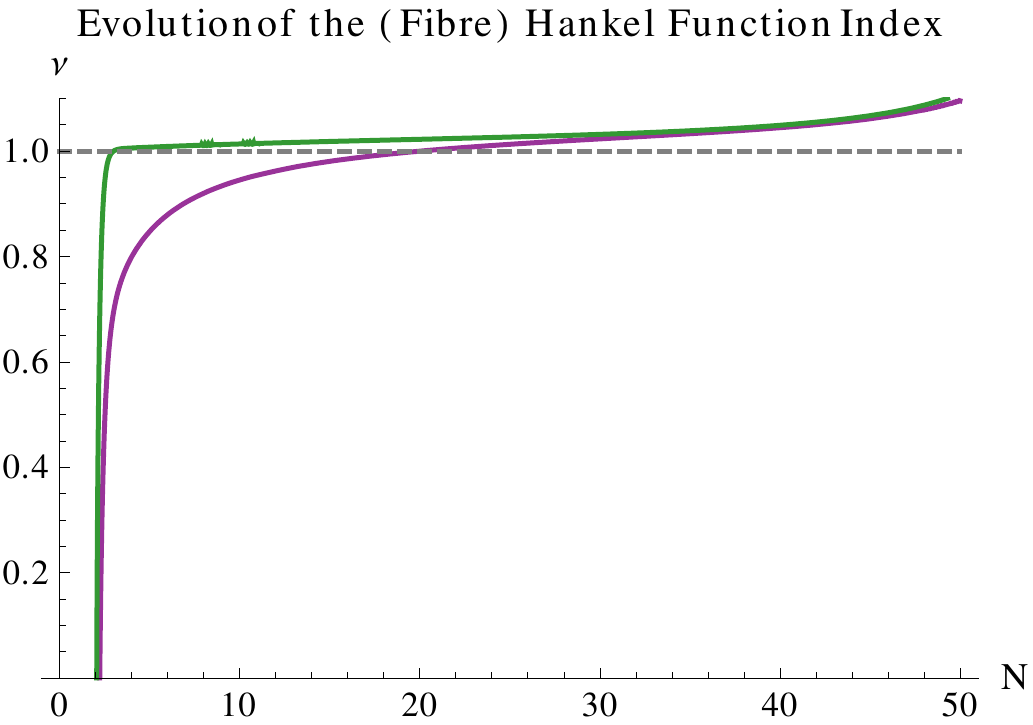}
\caption{Evolution of $\Xi = \phi^{\prime\prime}/\phi^{\prime}$, the deviation from the slow-roll condition \eqref{eqn:inexact}. Green and purple curves represent $n=2.1$ and $1$, respectively. The former shows a much stronger deviation from slow-roll, giving rise to a steeper transition as with the inflection point case. }\label{fig:FIchi}
\end{figure}

%


\subsection{The Onset of Inflation}

In \cite{Cicoli:2013oba,Pedro:2013pba} it was shown that Fibre inflation allows for a primordial origin for the low power at large scales anomaly in the CMB data. If the field starts above the plateau, a sufficiently rapid shift from a blue to a red spectrum of primordial density perturbations is required to fit the data. This amounts to a quick transition between a fast-rolling phase down the exponential wall and slow-rolling phase along the plateau. Given the attractor dynamics of the inflaton, the fast-roll phase occurs with
\begin{equation}\label{eqn:fibvel}\phi^{\prime}\approx \frac{V_{\phi}}{V}\approx -\frac{2(1+n)}{\sqrt{3}}.\end{equation}
If the magnitude of the RHS of \eqref{eqn:fibvel} is greater than $\sqrt{6}$, appealing to \eqref{eqn:FEQ} shows that the fast-roll solution becomes the attractor. The very flat plateau affords a sharp transition, qualitatively similar to the inflection point case, as communicated through Fig.~\ref{fig:FIchi}. Therefore, the arguments of the previous also carry through, giving rise to a similar angular power spectrum, \cite{Cicoli:2013oba}.


It should be emphasized that demanding a sufficiently step wall --- or large $n$ --- amounts to demanding a redshifted spectrum for the angular moments $\ell\gtrsim 50$. In short, it is fitting the spectral index to the data. Of course, a smaller value of $n$ is consistent so long as the initial value of $\phi$ remains close to the point of inflection at the start of the plateau. While this relaxes the tuning of $\delta$, to observe the power suppression from the onset of inflation, one must then explain why this is so. This is, in effect, the standard ``tuning'' problem of large-field inflation discussed above.

\section{Conclusion}

The Planck data favor inflation models with a redshifted scalar spectrum and a low tensor-to-scalar ratio. Two widely studied scenarios within particle physics and supergravity models are inflection point inflation and the Starobinsky model. We have reported on novel features of both scenarios, emphasizing the existence of an upper bound of $N_e$ induced by the couplings in both scenarios. We have argued that a similar mechanism is in play for both cases: degenerate critical points of the potential.

In both cases, the spectrum of perturbations transitions from blue to redshifted as the VEV rolls past a point of inflection in the potential. We have demonstrated that such transitions can be sufficiently sharp to allow for a finite number of e-foldings \textit{and} a redshifted spectrum. Therefore, both scenarios readily offer an explanation for the low power at large scales anomaly whose statistical significance --- while far from definitive --- has persisted in the Planck data.

\Acknowledgements
I am grateful to Kuver Sinha, Bhaskar Dutta and Michele Cicoli for collaboration on the work detailed here, also thanks are  owed to Joe Bramante, Eiichro Komatsu, Francisco Pedro, Nick Suntzeff and Alexander Westphal for useful discussions and perspectives.

\bibliographystyle{h-physrev.bst}

\bibliography{bib4cospa}

\end{document}